\documentclass[letterpaper, 10 pt, conference]{ieeeconf}  % Comment this line out if you need a4paper

\IEEEoverridecommandlockouts                              % This command is only needed if 
\usepackage{graphics} % for pdf, bitmapped graphics files
\usepackage{epsfig} % for postscript graphics files
\usepackage{mathptmx} % assumes new font selection scheme installed
\usepackage{times} % assumes new font selection scheme installed
\usepackage{amsmath} % assumes amsmath package installed
\usepackage{amssymb}  % assumes amsmath package installed
\let\labelindent\relax
\usepackage{enumitem}
\usepackage{url}
\usepackage[nolist]{acronym}
\newacro{vr}[VR]{Virtual Reality}
\newacro{hmd}[HMD]{Head-Mounted Display}
\newacro{ems}[EMS]{Electrical Muscle Stimulation}
\newacro{imu}[IMU]{Inertial Measurement Unit}
\newacro{bci}[BCI]{Brain-Computer Interface}
\newacro{vwg}[VWG]{Virtual World Generator}
\newacro{moo}[MOO]{Multi-Objective Optimization}
\newacro{dof}[DoF]{Degree of Freedom}
\newacro{ssq}[SSQ]{Simulator Sickness Questionnaire}
\newacro{ddr}[DDR]{Differential Drive Robot}
\newacro{fov}[FOV]{Field-of-View}
\newacro{ar}[AR]{Automatic Rotations}
\newacro{ur}[UR]{Unwound Rotations}
\newacro{osf}[OSF]{Open Science Foundation}
\newacro{vims}[VIMS]{Visually Induced Motion Sickness}
\newacro{ros}[ROS]{Robot Operating System}
\usepackage{flushend}

\usepackage{caption}
\usepackage{subcaption}

\usepackage{xcolor}

\title{\LARGE \bf
A Study of Preference and Comfort\\ for Users Immersed in a Telepresence Robot
}

\author{Adhi Widagdo, Markku Suomalainen, Basak Sakcak,  Katherine J. Mimnaugh,\\ Juho Kalliokoski, Alexis P. Chambers, Timo Ojala, and Steven M. LaValle% <-this % stops a space
\thanks{*This work was supported by a European Research Council Advanced Grant (ERC AdG, ILLUSIVE: Foundations of Perception Engineering, 101020977), Academy of Finland (projects PERCEPT 322637, CHiMP 342556), and Business Finland (project HUMOR 3656/31/2019).}% <-this % stops a space
\thanks{Authors are with Center of Ubiquitous
Computing, Faculty of Information Technology and Electrical Engineering,
University of Oulu, Finland. {\tt\small \{name.surname\}@oulu.fi}}%
}

\begin{document}

\maketitle

\thispagestyle{empty}
\pagestyle{empty}

%%%%%%%%%%%%%%%%%%%%%%%%%%%%%%%%%%%%%%%%%%%%%%%%%%%%%%%%%%%%%%%%%%%%%%%%%%%%%%%%
\begin{abstract}
In this paper, we show that unwinding the rotations of a user immersed in a telepresence robot is preferred and may increase the feeling of presence or\textit{} ``being there''. By immersive telepresence, we mean a scenario where a user wearing a head-mounted display embodies a mobile robot equipped with a 360\textdegree~camera in another location, such that the user can move the robot and communicate with people around it. By unwinding the rotations, the user never perceives rotational motion through the head-mounted display while staying stationary, avoiding sensory mismatch which causes a major part of VR sickness.
We performed a user study (N=32) on a Dolly mobile robot platform, mimicking an earlier similar study done in simulation. Unlike the simulated study, in this study there is no significant difference in the VR sickness suffered by the participants, or the condition they find more comfortable (unwinding or automatic rotations). However, participants still prefer the unwinding condition, and they judge it to render a stronger feeling of presence, a major piece in natural communication. We show that participants aboard a real telepresence robot perceive distances similarly suitable as in simulation, presenting further evidence on the applicability of VR as a research platform for robotics and human-robot interaction.

%Although robotic telepresence can be enhanced immersively using Virtual Reality, VR users commonly suffer from cybersickness.  The unwinding rotation is introduced as a challenge to improve comfort and reduce VR sickness. The approach includes the robot carrying a panoramic 360 degree camera, video audio streams, robot navigation deployed in remote locations, and immersive video playback in the head mounted display…..
\end{abstract}

%%%%%%%%%%%%%%%%%%%%%%%%%%%%%%%%%%%%%%%%%%%%%%%%%%%%%%%%%%%%%%%%%%%%%%%%%%%%%%%%
\section{INTRODUCTION}\label{sec:intro}
Telepresence robots are a rising trend in robotics, with commercial companies such as GoBe and Double marketing their robots as tools for hybrid meetings.
%While the popularity of telepresence robots has increased, and (GoBe and Double), many technical problems have occurred. 
However, these commercial robots do not make users feel like they really were at the robot's location, due to, for example, the inability to look around: %. Commercial products currently only have a little autonomy regarding the views they see because when the user gives control, it is possible that the other person who is facing the robot  is at a disadvantage. 
%It is also less effective since 
the robot should do a rotation to just explore the view. Additionally, this lack of immersiveness leads to a lack of presence, the feeling of "being there" \cite{slater_psi:2009}, which is essential for a user to communicate naturally with the people around the robot.  

\begin{figure}[t]
    \centering
    \includegraphics[width=0.6\columnwidth]{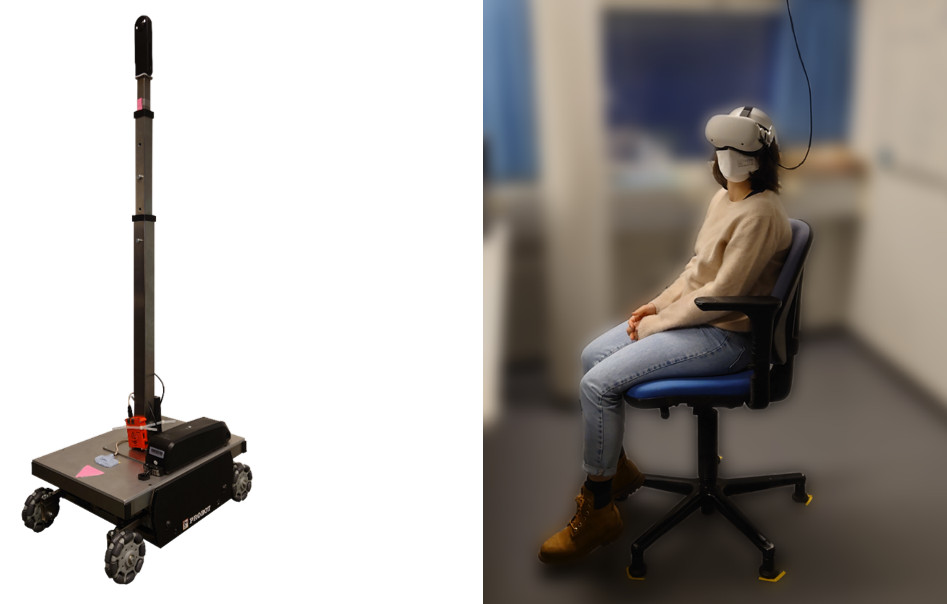}
    \caption{Robotic telepresence: a user wearing an HMD can embody a mobile robot equipped with a 360\textdegree~camera. Left: Dolly mobile robot platform and a 360\textdegree~camera on a boom, used in this study. Right: the physical setup of our user study.}
    \label{fig:telepresence}
    \vspace{-0.55cm}
 \end{figure}

Virtual reality has emerged in the robotics field to complement the trend of hybrid meetings, which have recently increased significantly. The advantage offered by integrated robotics-VR, such as shown in Fig.~\ref{fig:telepresence}, is the increased immersion of the \ac{hmd}, which can make the user really feel present. With the use of a 360\textdegree~camera capable of capturing omnidirectional videos, instead of a standard camera used in the current commercial telepresence robots, %innovative applications can be implemented on telepresence robots. So that 
the experience %felt in VR 
will be more immersive. %Telepresence robot also looks promising whether it has the ability to explore certain areas both automatically or remotely.
%Immersive robotic telepresence can be achieved by using a 360\textdegree~camera as a video capture that will be viewed by the user on a \ac{hmd}. 

However, using an \ac{hmd} can create side effects, namely VR sickness, with symptoms such as dizziness, headache, eye strain, blurred vision, vomiting, and nausea  \cite{laviola2000discussion}. This is caused by a sensory mismatch between vision and the vestibular organ; when the user's eye informs the brain about acceleration of the body, but the acceleration sensor in the form of the utricle and saccule in the otolith organ does not, and the result is a conflict with the information received by the brain \cite{LaValle_bookVR}. In addition, feeling the immersiveness can also cause discomfort if objects are too close, or if rotations of the robot are surprising, issues that are not present in a conventional telepresence scenario \cite{suomalainen2021comfort,mimnaugh2021analysis}.

In this paper, we present a user study on immersive telepresence where participants embodied a real mobile robot that was equipped with a 360\textdegree~camera and traversed a predefined path at a university campus. We explore \textit{unwinding the rotations} of the user's viewpoint, which has been shown to reduce VR sickness in a study based on a simulation \cite{suomalainen2022unwinding}. This method negates the rotation of the user's view relative to the robot, so that the direction of the view is determined solely by the user. %This also does not affect the direction of the robot's motion, so the robot has more freedom to move in the path and a safe distance from the wall.
%The advantage is to reduce the lateral optical flow, which causes more sickness than optical flow when moving forward, in the user's view when the robot rotates. 
We test the same three hypotheses that were confirmed about unwinding rotations in the simulation study. However, this time only the user preference of unwinding the rotations is confirmed; there is no statistically significant difference in user comfort or \ac{ssq} measures, a commonly used questionnaire for VR sickness. We hypothesize, based on answers to open-ended questions, that these different results are more likely due to simply a different sample than other issues such as video quality or the vibrations of the real mobile robot missing from the simulation study, even though the latter cannot be ruled out. These findings call for further studies with more participants with both simulated and real mobile robots. Additionally, we observe participants feeling present enough that 11 out of 32 respond (with either words or gestures) to people in the video waving or talking to the robot, even though the participants are told it is a recording; there is also a perceived difference in how present participants feel between the unwinding and coupled rotations conditions. Finally, we confirm that users perceive the distances to passing objects and the speed of the real robot similarly as in the simulation study, giving further justification to study telepresence and HRI with simulations in the future.  %Based on the sensory mismatch theory, we believe that this filter method will further reduce the feeling of sickness. 
 
%The overarching purpose of this research is to explore the functional effectiveness and user comfort of immersive telepresence systems. Functional effectiveness refers to how well a system achieves its targeted goals such as the ability to compensate for the lateral optical flow due to robot rotation and exploration of remote places. User comfort refers to how much (dis)comfort, such as fatigue or nausea, results from using the system. We also explore user behavior and preferences regarding self-rotation and viewpoint selection aboard a moving mobile robot. 

%e34444444444
 
%\vspace{-1mm}

\section{RELATED WORK}
Most of the previous research investigating teleoperation and telepresence, the idea of embodying a robot equipped with a camera, used standard screen-based technologies, before consumer grade VR headsets and 5G networks \cite{kheddar2014virtual}. There is an unprecedented opportunity to leverage these technologies to bring telepresence to a new level \cite{bastug2017toward} \cite{aykut2018delay}. %In general, telepresence uses a robot equipped with a camera. %This is intended in a way users can know and feel the conditions in which the robot is moved.
The key improvement that an \ac{hmd} can bring to telepresence is increased immersion, which in turn can lead to greater feelings of presence \cite{skarbez2017survey}. Standard telepresence robots equipped with 2D screens provide similar improvement over conventional videoconferencing by enhancing interaction and navigation tasks for remote users \cite{rae2014bodies}; however, the feeling is still closer to video conferencing than really being there. %Visual Information significantly helps remote users of telepresence robots for doing interaction and navigation tasks \cite{rae2015framework}.
%Telepresence is also equipped with a monitor on the robot's body that functions for social interaction \cite{kristoffersson2013review}, some of which are very useful for video conferencing with colleagues \cite{tang2012social} and engagement in collaborative work \cite{lee2011now, venolia2010embodied}
Use cases for telepresence robots are, for example, distance education \cite{botev2021immersive}, telemedical consultation \cite{carranza2018akibot} and all kinds of remote collaborative work \cite{lee2011now, venolia2010embodied}. %However, some of the telepresence systems \cite{johnson2015can} have limitations for viewing behind the robot in which it requires the robot to do maneuvering. Also the situational awareness is limited which affects the presence. 
 
While a lack of presence can reduce task performance \cite{stoll2018wait}, wider field of view improves awareness and immersion for the user of telepresence robots as well as tasks efficiency \cite{johnson2015can}. Using an HMD in telepresence for increased immersion has been reported to yield better results in many application domains including education \cite{botev2021immersive}, collaboration through conversation \cite{du2020human}, scenarios for earthquakes \cite{negrello2018humanoids}, and advanced navigation for robots \cite{baker2020towards}. However, for immersive \ac{hmd}-based telepresence to reach a wider audience, the comfort of the HMD user and reducing VR sickness needs to be studied further.
 
VR sickness typically occurs due to a mismatch between the real-world and perceived accelerations in an HMD \cite{rebenitsch2016review}. Also, issues with hardware and content such as repeated motions often lead to nausea \cite{keshavarz2014visually}. %Prior literature has reported that wider field of view may cause motion sickness \cite{smyth2000indirect}.
Several approaches have been introduced for reducing motion sickness, such as minimizing the optical flow to slow down motions \cite{laviola2000discussion}, adding a rest frame to the user's view \cite{cao2018visually}, or using a narrow field of view \cite{Teixeira_Palmisano_2020}. However, an important aspect is to consider criteria based on human perception \cite{urvoy2013visual}. In multimedia, quality of experience (QoE) measures have been developed as system performance criteria \cite{keshavarz2014visually}. Therefore, it is worthwhile to explore the design of robot motions that would reduce accelerations mismatches, and how such motions could be realized for robots used in immersive telepresence. %Other factors such as spatial frequency, contrast, brightness, distance to visual features, and field of view must also be considered and compensated for without substantially reducing the sense of presence in a remote or virtual environment.

Besides telepresence, VR is becoming a popular technology in robotics research in general, especially in human-robot interaction, as VR enables cost-effective simulations of complex real world use cases. However, the applicability of VR simulations is an open question; how much can the results received in VR be generalized to the real world? There are a growing number of studies on the topic, such as the comparison of proxemics between robots and humans in VR and in the real world \cite{li2019comparing}, robot navigation among people in VR \cite{grzeskowiak2020toward}, and designing robots for human-robot interaction in VR \cite{zamfirescu2021fake}. In this paper, we continue this trend by exploring whether participants aboard a real telepresence robot perceive objects or people similarly or differently (too close or too far) to a previous simulation study.

\section{Unwinding Rotations}
In \cite{suomalainen2022unwinding}, we introduced an approach for decoupling the user's viewpoint from the rotations of the robot, which we called \textit{unwinding rotations}. We present our approach applicable for mobile robots here for completeness; details on applying the method to robots with more complex motions than on a plane can be found from \cite{suomalainen2022unwinding}.

We consider a mobile robot moving in a two-dimensional plane. The discrete-time kinematic model of the robot is given as
\begin{equation}
    \begin{split}
    x_{k+1} & = x_k + v_k \cos(\theta_k)dt\\ 
    y_{k+1}& = y_k + v_k \sin(\theta_k)dt\\ 
    \theta_{k+1}&=\theta_k+w_kdt,
    \end{split}
\end{equation}
in which $(x_k,y_k)$ is the position and $\theta_k$ is the orientation of the robot at time instance $kdt$ with respect to an absolute reference frame and $v_k,w_k$ are the control inputs corresponding to the linear and angular velocity with respect to a robot-fixed reference frame. We assume that the control input is kept constant for time window $t \in [kdt, (k+1)dt]$. 

Suppose that the robot orientation $\theta_k$ can be estimated and denote its estimate by $\hat{\theta}_k$. Let $R(\hat{\theta}_k) \in SO(2)$ be the rotation matrix corresponding to the robot orientation. 
Then, we can define unwinding rotations as rotating the camera frame such that any point $p_c \in \mathbb{R}^3$ represented in the camera frame is related to the point $p'_c$, whose coordinates are expressed in the rotated camera frame, through
\begin{equation}
   p'_c=\begin{bmatrix}
R(\hat{\theta}_k)^T & 0_{2\times1}\\
0_{1\times2} & 1
\end{bmatrix}p_c.
\label{eq:unwind}
\end{equation}
This operation negates the rotation of the viewpoint caused by the robot rotation. Thus, the users need to rotate themselves in order to face the direction that the robot is facing or face the same direction in the virtual environment for the whole motion.

%This is illustrated in Fig.~\ref{fig:unwinding}: the robot is moving, either autonomously or controlled by a user directly, and will take a right at the corner. Suppose the user's viewpoint before the turn is towards the wall at the end of the first corridor (view in Fig.~\ref{fig:unwinding}a). If we are not unwinding the   rotation that the camera frame undergoes (Fig.~\ref{fig:unwinding}f) due to the robot motion, user's viewpoint will change towards the robot heading as the robot rotates, even if she has not moved her head (Fig.\ref{fig:unwinding}c). The proposed unwinding rotations method ensures that, by rotating the camera frame (Fig.~\ref{fig:unwinding}e), after the robot has made the turn, user's viewpoint will no longer be towards the robot's direction of motion but it will be towards the wall unless she physically moves her head (Fig.\ref{fig:unwinding}b). 

\section{Conditions and Hypotheses}
In the \textbf{unwound rotations (UR)} condition, the user’s viewpoint does not change when the robot changes direction. In the \textbf{coupled rotations (CR)} condition, the user’s viewpoint rotates when the robot rotates. %Additionally, both rotations complement the user’s rotation.

We pre-registed the following three hypotheses in the Open Science Foundation (OSF, \url{https://osf.io/tw7ef}). They are identical to the confirmed hypotheses of a previous study with a simulated robot in a virtual environment \cite{suomalainen2022unwinding} (\url{https://osf.io/eks6t}), as the goal of this study is to test the hypotheses with a real robot.  

\begin{enumerate}[label=\textbf{H\arabic*:}]
    \item %H1: 
    Less VR sickness in UR condition as indicated by lower total weighted \ac{ssq} score. 
    \item The UR condition is more comfortable as indicated by asking directly which was more comfortable (forced choice).
    \item The UR condition is preferred as indicated by asking directly which the user preferred (forced choice).
\end{enumerate}

According to VR sickness theory, that a mismatch between vestibular system and vision induces sickness, we predicted in \textbf{H1} that there will be less sensory conflict if the user performs their own physical rotations. \textbf{H2} and \textbf{H3} were under the general assumption that people avoid sickness, hence UR is most likely to be preferred and comfortable.

\section{Methodology}\label{sec:methods}
%In this section we explain the hardware and procedures used in the user study. The camera system and filtering methods for the video stream are explained in Section~\ref{sec:video}, and in Section~\ref{sec:dolly} we describe the mobile robot on which the camera is attached. In the rest of Section~\ref{sec:methods} we give details on the user study, such as participants, procedure, and the metrics used. 

%There are four kinds of data used in this project. The first data is in the form of VR content presented at HMD in the form of panoramic video streams and applications on Unity. The second data is obtained from the sensor output in the form of IMU, camera, laser, and odometer. The third data is obtained from computational motion planning for robot navigation. The fourth data will be collected from experiments conducted on  human subjects using HMD and questionnaires in the form of variations of SSQ \cite{kennedy1993simulator} and SUS \cite{slater1994depth, slater1997framework}. The data will be used to provide solutions to hypotheses about increasing comfort for VR users and getting the desired level of telepresence.

 \begin{figure}[t]
     \centering
    \includegraphics[trim=2 3 2 2,clip,width=\columnwidth]{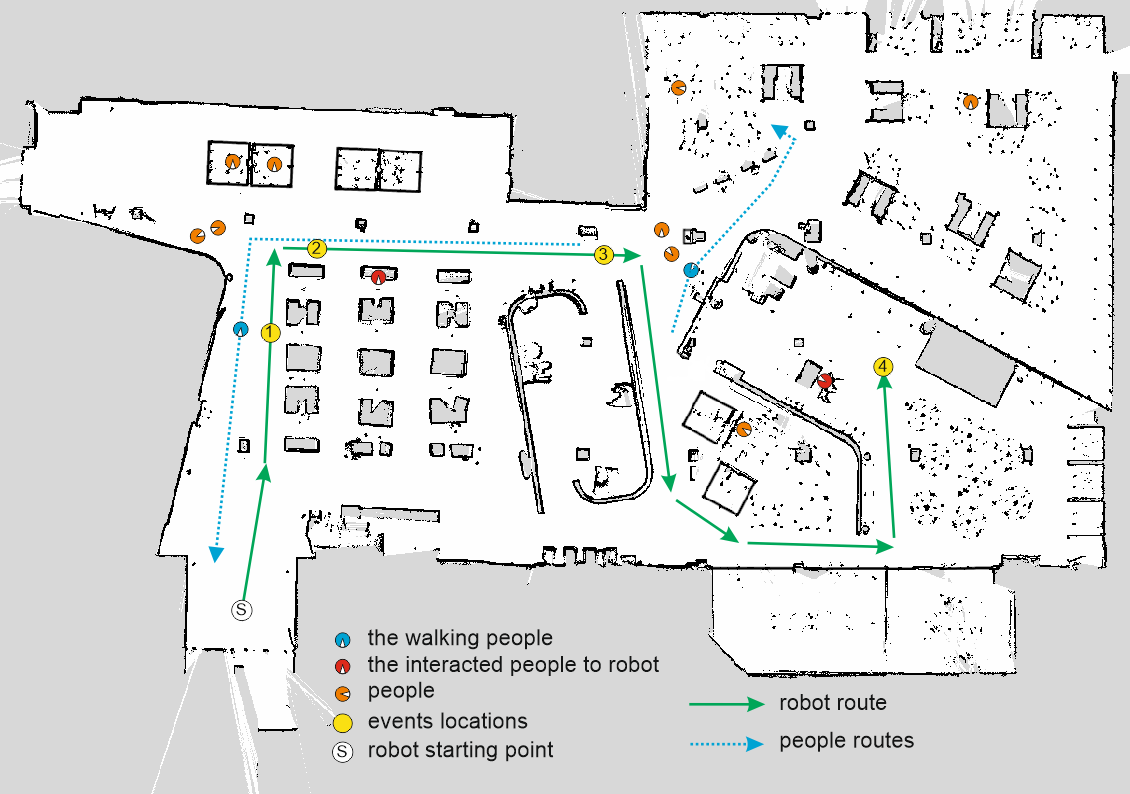}
     \caption{A top-down view of the environment tested trajectories.}
     \label{fig:tellus_map} 
     \vspace{-3mm}
 \end{figure}

\subsection{Mobile Robot}
\label{sec:dolly}
We used a wheeled mobile robot Probot Dolly (Fig.~\ref{fig:telepresence}, left) equipped with a 360\textdegree~camera mounted on a vertical boom. The dimensions of the robot are 60 (L) x 40 (W) x 20 (H) cm. The robot uses 2 active wheels and 4 omni wheels so that the robot's movement is based on a differential drive mobile robot (DDMR). The robot is equipped with a laser scanner and an odometer to move according to the direction of the navigation, which is controlled by a motion planning algorithm. The robot uses a client-server software architecture where a laptop hosts Robot Operating System (ROS2) galactic as the main operating system and as the server. The main control board of the robot is a Raspberry Pi 4 and it acts as a client. The client and the server communicate over a WiFi connection using the TCP/IP protocol. A laser connected to Ethernet port of the main control board is used for mapping (SLAM) and obstacle avoidance. The actuator integrated with the HAL sensor is driven by the VESC controller which is connected to the CAN-bus so that it can function as an odometer.

\begin{figure*}%
  \centering
    \begin{subfigure}{0.44\columnwidth}
        \includegraphics[width=\columnwidth]{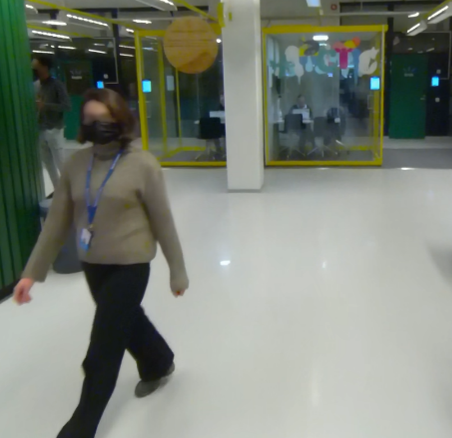}
        \caption{}\label{fig:walking_lady}
    \end{subfigure}\hfill%
    \begin{subfigure}{0.39\columnwidth}
        \includegraphics[width=\columnwidth]{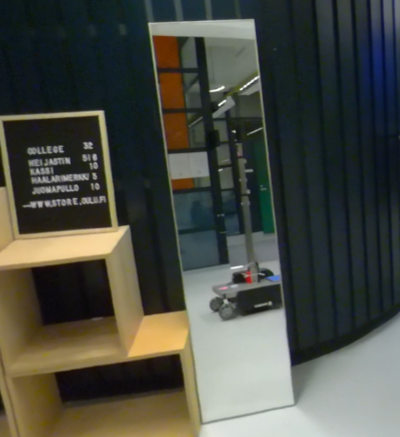}
         \caption{}\label{fig:robot_reflection}
    \end{subfigure}\hfill%
    \begin{subfigure}{0.73\columnwidth}
      \includegraphics[width=\columnwidth]{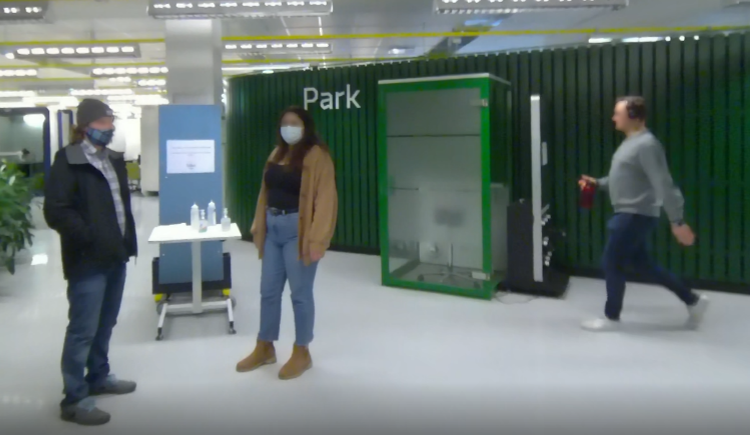}
    \caption{}\label{fig:people_standing}
    \end{subfigure}\hfill
    \begin{subfigure}{0.44\columnwidth}
      \includegraphics[width=\columnwidth]{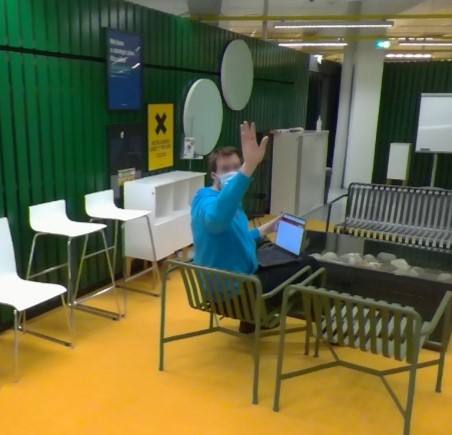}
    \caption{}\label{fig:person_greeting}
    \end{subfigure}%
\caption{Screenshots from the environment from the user’s view:  (a) a person passing the robot; (b) A VR user watching their own body through a mirror's reflection; (c) People standing in the midway; (d) a person greeting and asking a VR user. }
\label{fig:screenshots}
\vspace{-0.5cm}
\end{figure*}

In ROS2, we used waypoint based navigation and dynamic window (DWB) approach as controllers to follow the path. The robot moved inside the building of the University of Oulu campus with a maximum speed of $0.7~m/s$, a maximum rotating speed of $0.7~rad/s$ and a maximum rotational acceleration of $2.5~rad/s^{2}$. The length of the path traversed by the robot is 61 meters, the route is shown in Fig.~\ref{fig:tellus_map}; screenshots from the video are presented in Fig.~\ref{fig:screenshots}. The minimum distance to the obstacles is $0.95~m$ and the minimum distance to the people is $1.2~m$.

%Additional sensors in the form of IMU and a panoramic camera use SSH protocol and RTSP respectively. 

\subsection{360\textdegree~Video}
\label{sec:video}
We used a commercial Kandao QooCam 8K camera to record 360\textdegree~video as the robot moves in the environment. The camera was mounted on a vertical boom so that the center of the camera lenses was 155 cm above floor level. The camera captured a video of 7680x3840 pixels in resolution at 30fps with H.265 encoding. The duration of the video was $2~minutes$ $6~seconds$, during which the robot rotated a total of $648.7$\textdegree. This camera has a built-in IMU sensor, making it easy to estimate $\hat{\theta}_k$ for (2). However, for earlier tests we also implemented unwinding rotations on a camera which did not have a built-in IMU; for such cameras the key point is the synchronization of the IMU and camera data. We also note that in these studies the unwinding rotations was performed as post-processing; adapting the method for online processing is part of future work once we establish that unwinding rotations is favored by the users.

The original video had equirectangular projection for the mapping between 2D textures and 3D direction vectors for tilling the surface of a sphere \cite{zucker2018cube}. However, such a sphere suffers from seam-like artifacts commonly occurring around the poles \cite{azevedo2019visual}. We argue that seeing such seam-like artifacts would decrease immersiveness. To remove the seams, we constructed a uniform spherical grid that is originally transformed from a cube (cube-sphere) \cite{rocsca2011uniform}. First, video textures from the equirectangular mapping are reconstructed into the cube mapping by shaders at the GPU level. Then, the video’s UV coordinates are uniformized with the cube-sphere’s UV mapping, using coordinates generated by the cube-sphere mesh generator, so that the image will be fully projected on the geometry. We render the video under the Unity engine on a desktop PC furnished with an Intel Xeon 3.80 GHz CPU, 32 GB RAM and NVidia Quadro RTX6000 GPU, and transmit the video to an Oculus Quest 2 HMD via a link cable.

%\todo{TODO: Explain Cube-Sphere better: what is it? A parametric geometric shape such that $\alpha=1$ it is a sphere and $\alpha=0$ it is cube. Add reference to Cube-Sphere whatever..\cite{rocsca2011uniform,zucker2018cube} }

\subsection{Procedure}
We counterbalanced both the order of the videos (which video is played first) and gender by presenting the two videos of the UR and CR conditions equally for both men and women as the first video. Each participant was asked to read the privacy notice and study information, and to sign the consent form. Before starting the experiment, the researcher did pre-screening such as asking whether the participant has a headache or feels nauseous (rescheduling was an option for a participant if they felt unwell). After that, they were asked to sit and adjust the chair’s height on the swivel chair so that they could rotate around easily (see Fig.~\ref{fig:telepresence}, right). Next, the researcher gave the instructions and told the participant about the brief experiment information, including how to put on the HMD.  Afterwards, the researcher started the video, calibrated the participant position, and asked them to rotate around while wearing the HMD to make sure they understand they can easily rotate with it; they were however not prompted to rotate themselves during the video. After the participant confirmed that they are ready to start, then the video of the robot moving started playing. During the video, the participant’s head orientation was recorded. After the video finished, participants were asked to take off the headset and fill out the questionnaire regarding their experience. The same procedure was applied for the second video. The participants were compensated with a 20€ Amazon gift card. We used recommended precautions for COVID-19 during the experiment such as wearing a mask, keeping a safe distance from the subject and sanitizing the devices (including the environment). 

\subsection{Measures}

In terms of questionnaires, we measured sickness with the SSQ \cite{kennedy1993simulator} which presented 16 possible sickness symptoms, rated on a scale of none (0) to severe (3). The SSQ total score and subscales are calculated by weighting the answers for a maximum score of 236. We used the SUS questionnaire \cite{slater1994depth,usoh:2000} to attempt measuring their feeling of presence. The experience questionnaires also included Likert-scale (ranging 1-7) questions and  open-ended questions for reasoning in some choices. After the second video, there were more questions in the shape of forced-choice questions to compare the two videos in regards to choosing the more comfortable, preferred, and more intuitive condition for the participant. Furthermore, we also asked questions (in Likert-scale) about the feeling towards distance to walls, distance to humans, and linear speed of the robot. In the end, the VR and gaming experience, and demographics were also queried in the questionnaire.  

\subsection{Participants}
We collected the data from 32 participants similarly to the prior study \cite{suomalainen2022unwinding} consisting of equal numbers of both males and females. Their age ranged from 20 to 33 years old and they reported having normal or corrected-to-normal vision. There were no reports of color blindness. The responses of how often they use VR systems were as such: $18.8\%$ said never, $53.1\%$ once or just a couple of times ever, $9.4\%$ once or twice a year, $15.6\%$ once or twice a month, and $3.1\%$ once or twice a week. Meanwhile, they also responded of how often they play computer games: $15.6\%$ never, $25\%$ once or just a couple of times ever, $25\%$ once or twice a year, $9.4\%$ once or twice a month, $12.5\%$ once or twice a week, $9.4\%$ several times a week, and $3.1\%$ everyday.

\section{RESULTS}
We ran data analysis in SPSS with the Wilcoxon signed-rank test to find whether there was a difference in the total weighted SSQ score under two different video conditions (UR and CR). An exact binomial test with exact Clopper-Pearson $95\%$  CI was used to determine if a greater proportion of participants were more comfortable and preferred one of the videos when shown UR compared to the CR conditions.

The SUS score was calculated as the number of SUS questionnaire items given a 6 or 7 rating by a participant, resulting in a score that could range from 0 to 6 for each participant \cite{slater1994depth,usoh:2000}. We performed a Wilcoxon
signed-rank test (one-sided) to test for greater SUS scores 
%in the body condition than in the without body condition.

\subsection{Confirmatory Results}
%One paragraph: SSQ Wilcoxon signed rank

%One paragraph: Forced-choice comfort, Exact binomial, one-sided

%One paragraph: Forced-choice preference, Exact binomial, one-sided

A Wilcoxon signed-ranks test showed that UR did not elicit a statistically significant median decrease in the total weighted SSQ score ($Z= -0.144, p= .886$) with median $14.96$ and $16.83$ for the CR.

\begin{figure}
    \centering
    \includegraphics[trim=2 3 2 2,clip,width=0.9\columnwidth]{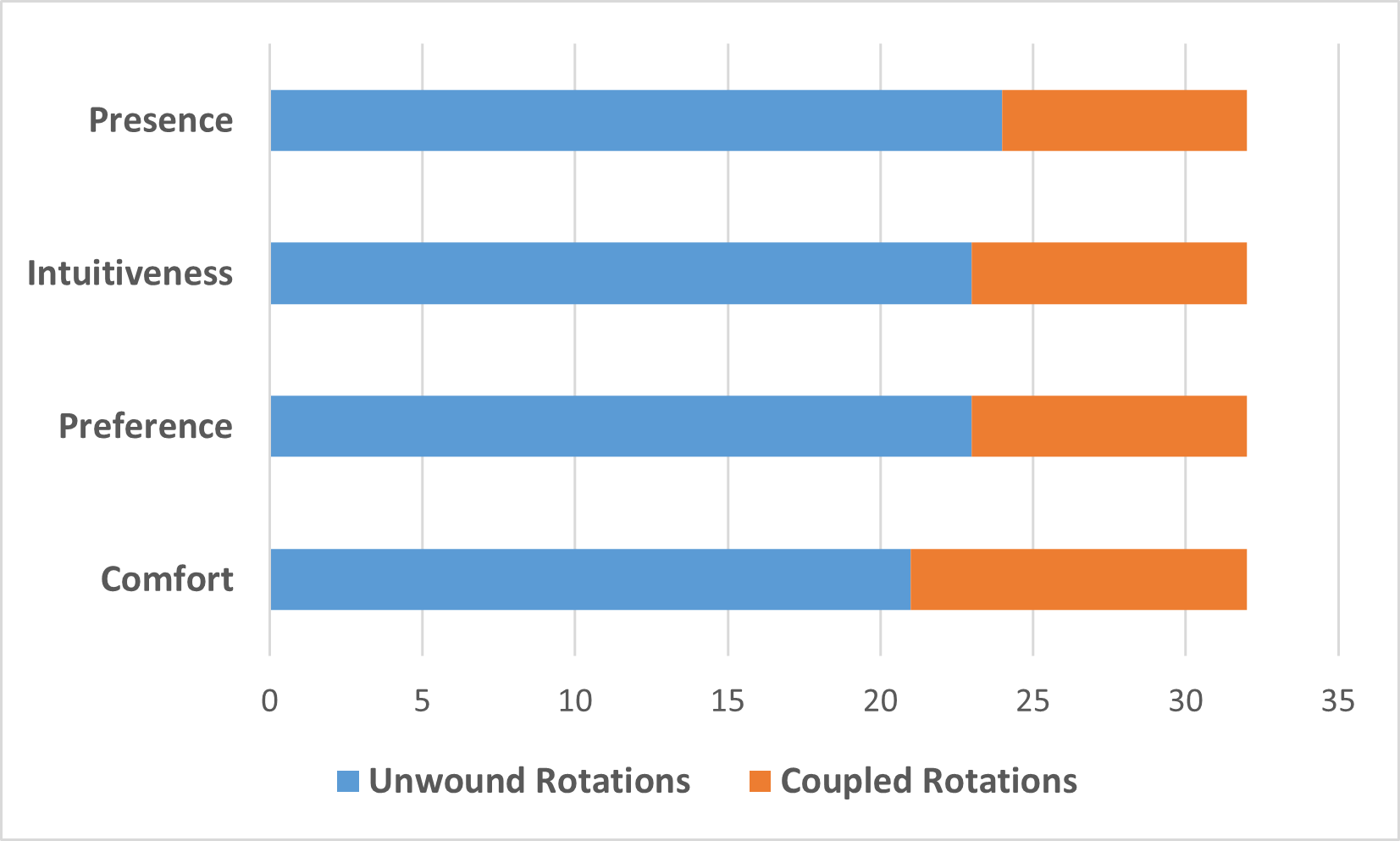}
    \caption{The distributions of responses to the questions regarding comfort, preference, intuitiveness, and presence.}
    \label{fig:force_choice_accumulation}
    \vspace{-0.35cm}
\end{figure}

While 21 out of 32 participants felt more comfortable in UR ($65,6\%$) than the CR, 23 participants preferred UR ($71,9\%$) when they responded to the forced-choice questions: “Which of the two videos is more comfortable?” and “Which of the two videos do you prefer?” (see Fig.\ref{fig:force_choice_accumulation}). A binomial test with exact Clopper-Pearson found that the condition UR was not significantly more comfortable, had  $95\%$ CI of $46,8\%$ to $81,4\%, p= .112$ and the preference in UR was statistically significant, had $95\%$ CI of $53.3\%$ to $86.3\%, p= .022$.

\subsection{Exploratory Results}

\subsubsection{Quantitative Data}
A Wilcoxon signed-rank test indicated no statistically significant difference in the Likert-scale comfort ratings $Z = -1.082, p= .279$ with $Mean = 5.03$ for UR and ($Mean = 4.78$) for CR.

In the SUS scores, UR ($Mean=4.96$) and CR ($Mean=4.84$) did not have a statistically significant difference in the sense of being in the remote environment, $Z= -.510, p = .610$.

%Interestingly, the UR ($Mean = 4.7$) had more time significantly that the participants thought the remote environment was real $Z = -2.716, p = .007$, (CR $Mean = 4.03$). 
%Also, UR ($Mean = 4.43$) showed a no significant difference in the state of being in a remote environment $Z= -1.664, p= .096$, (CR $Mean= 4.03$). 
%Moreover, other aspects in the SUS questionnaires resulted as follows: 
%“When you think back to the experience, do you think of the remote environment more as watching a movie or more as someplace that you visited?”, the UR ($Mean = 4.5$) and CR ($Mean=4.25$) didn’t have significant difference $Z = -.774, p= .439$
%“During the time of the experience, which was strongest on the whole, your sense of being in the remote environment, or of sitting at your original position?”, the UR ($Mean = 4.53$) had no significant increase rather than the CR($Mean = 4.18)  Z= -1.023, p= .306$
%“Consider your memory of being in a remote environment. How similar in terms of the structure of the memory is this to the structure of the memory of other places you have been today?”, both the UR and CR were similar and had no significant difference ($Mean=4.718) Z= -.026, p= .979$.

In addition to the pre-registered hypotheses, we asked two further forced-choice questions: “Which of the two experiences feels more intuitive for you?” and “Thinking back to both of the experiences, which one gave a better sense of being in the robot's location?”.  A binomial test with exact Clopper-Pearson showed that the UR was significantly more intuitive, $95\%$ CI of $53,3\%$ to $86.3\%, p= .022$ and elicited significantly better presence, $95\%$ CI of $56,6\%$ to $88.5\%, p= .008$ (see Fig.~\ref{fig:force_choice_accumulation}).

We asked Likert-scale questions about how the participants perceived the distance to walls, distance to humans and the speed during moving with the robot. The results shown in Table~\ref{tab:attributes} indicated that the speed and distance to people and objects were considered appropriate on a 7-point Likert scale.

%\todo{Add distance to objects and humans, and speed, means and variances to the likert scale questions}

\begin{table}
\centering
\begin{tabular}{l|l|l|l|l}
                         & \textbf{Value} & \textbf{Median} & \textbf{Variance} &  \\ \cline{1-4}
\textbf{Distance to walls}  & min. 0.95~m      & 4                     &     1.29               &  \\ \cline{1-4}
\textbf{Distance to people} & min. 1.2~m        & 4                     &         1.19           &  \\ \cline{1-4}
\textbf{Speed}           & max. 0.7~m/s      & 4                     &         0.547           &  \\ \cline{1-4}
\end{tabular}
\caption{\small{The values of certain attributes of the path, and the median and variance of 7-point Likert scale subjective opinions (1 meaning too small/too close).}}
\label{tab:attributes}
\vspace{-0.5cm}
\end{table}

%SUS: Follow what's above
%Forced-choice presence. Exact binomial, two-sided. 

%Distances to people and objects. Recover from ROSbag the closest distances to people and objects. 

%Either analyze the carryover effects for SSQ, or just cite previous unwinding and Mexico 1.2. Same for gender (IF we want to run it, use 30 people). 

%Likert-scale comfort.

%Head-motion direction, as in previous Unwinding. Maybe something like a heatmap, but as a kinda last thing to do. 

%Intuitiveness. Exact binomial. 

\subsubsection{Qualitative Data}
\begin{figure}[t!]
    \centering
    \includegraphics[trim=2 3 2 2,clip,width=0.9\columnwidth]{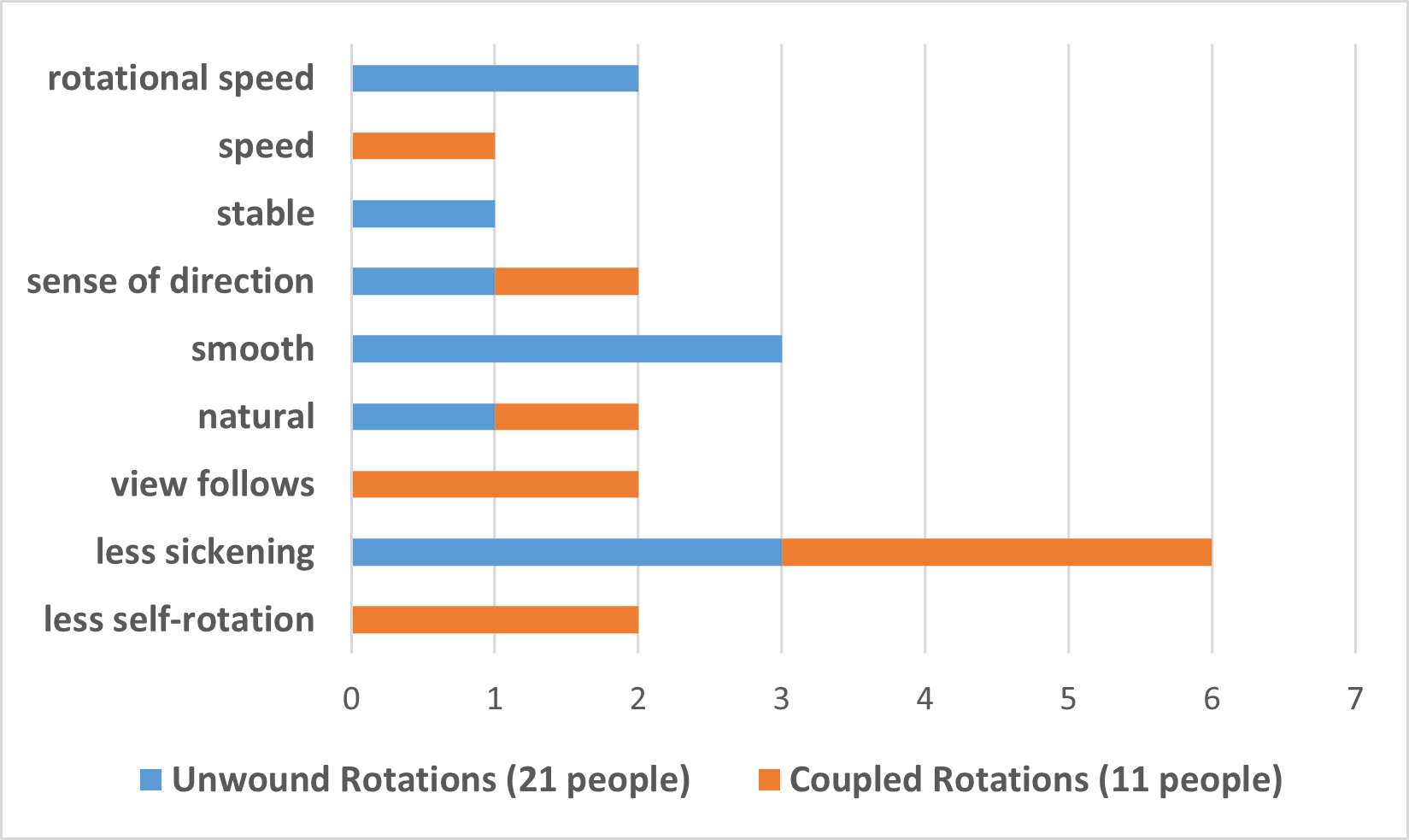}
    \caption{The frequently found codes from the question ``Please explain why that video is more comfortable” which was asked after the participant watched both videos.}
    \label{fig:comf_oe_responses}
    \vspace{-0.5cm}
\end{figure}

We analyzed the open-ended data using the inductive approach on thematic analysis \cite{Patton2005qualitative}. The accumulated responses for the open-ended question about why participants found either video more comfortable can be seen in Fig.~\ref{fig:comf_oe_responses} in which the most frequent keywords towards open-ended questions of UR as chosen comfortable video were as follows:  smooth (3) (for example, \textit{“The movement was smoother, both the straight-going parts, and the turning parts”}), less sickening (3)(\textit{“I did not feel nauseous at all after the first video, after the second one I felt a little nauseous”}), and natural (2) (\textit{“Moving with the robot felt more natural”}). 

On the contrary, the participants who chose the CR commented: 
less sickening (3) (\textit{``more static, so less motion sickness. Second video changing position/ changing gaze direction caused discomfort even though it was more engaging”}); 
less self-rotation (2)  (\textit{“The comfortable is quite the same, but I choose the first one since I dont need to turn the chair. Also, the second one can provide a reversed walk that is not natural for the speed”}); and
view follows (2) (\textit{“I did not require to turn to see where the robot was going”}).

\begin{figure}
    \centering
    \includegraphics[trim=2 3 2 2,clip,width=0.9\columnwidth]{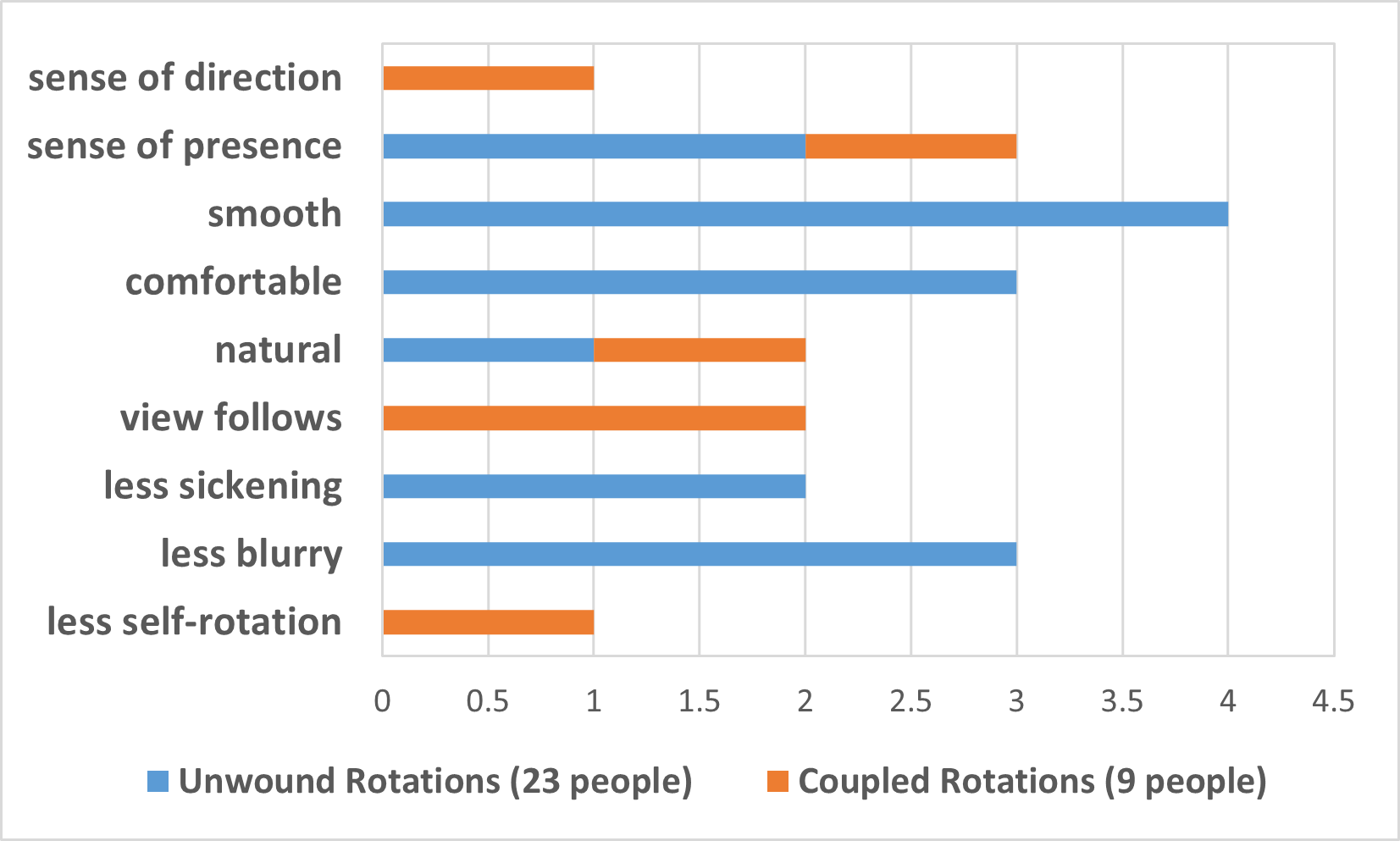}
    \caption{Frequently found codes for question ``Please explain why you prefer that video”.}
    \label{fig:pref_oe_responses}
    \vspace{-0.55cm}
\end{figure}

The open-ended question on the video preference was commented, shown in Fig.~\ref{fig:pref_oe_responses}: In UR there was 
smooth (4), \textit{``It was much smoother, especially the turning. I did not fully appreciate the first video, until after I watched the second one”}; 
control over viewpoint (5), \textit{``maybe more engagement with what I was seeing because I had to move to change view”}; comfortable (3), \textit{“It didn't feel uncomfortable at all and I could relax”}, natural (3), \textit{”it is quite real and comfortable”}; 
less blurry (3), \textit{``The video seemed clearer, and the view was less blurry even while robot was making turns”};
less sickening (2), \textit{``It didn't feel uncomfortable at all and I could relax”}; and 
sense of presence (2), \textit{``I think the first video gave me a better sense of actually being in that place, but I don't know if that's specifically because of a difference between the videos themselves or it is due to that being my first contact with the video”}. 

In contrast, the CR had comments such as sense of presence (1), \textit{``It feel [sic] like the two videos were honestly the same, but I chose the first one here, because it made me believe I was actually there more (as in I didn't think how I was sitting in the chair and so on)}” and  
view follows(2), \textit{``The first video was less disorienting and rather easier to maneuver in the space. The second video felt as if I was being pushed and not in line with my natural progression (of movement)”}.

During the experiments, we observed the participants’ reactions. We created the events (interaction to the robot) in the video, shown  in Fig.~\ref{fig:tellus_map}. The first event was when a lady walked past the robot, secondly a man waved his hand to the robot, thirdly the talking people standing in the middle of the path and a man crossed in front of the robot, and lastly a man greeted, waved hand and asked towards the robot. Surprisingly, there were some participants who responded back, especially towards the second and fourth events; in total 11 of the 32 participants responded to either of both people in the video communicating with them, either with gesture or even by speaking. There were also participants who answered the greeting at the last events.

\section{DISCUSSION}
From the pre-registered hypotheses confirmed in a simulation study \cite{suomalainen2022unwinding}, only \textbf{H3}, namely that users would prefer the unwinding, held: there was no statistically significant difference in either the SSQ scores or the forced-choice question of comfort. This is surprising, since there is not supposed to be a lot of difference between real 360\textdegree~video and a simulated one when it comes to VR sickness. It is possible that the use of the real robot brings in other causes of VR sickness, such as vibrations; however, this was not found in the open-ended answers, since the word "smooth" seemed most often to refer to the turns. Moreover, the mean SSQ value ($16.83$) for CR was exactly the same as in the simulation study, but the UR SSQ value was higher than in the simulation study ($14.96$ and $9.35$). As VR sickness is typically caused by multiple sources \cite{laviola2000discussion}, it is likely that increased vibrations would also have increased the SSQ scores of the CR conditions. We also note that there was a difference in the total rotation made by the robot: whereas in the simulation the robot rotated a total of 438\textdegree~, in this study the total amount of rotation was 648.7\textdegree. This was mainly due to the small corrections a robot needs to constantly make while moving; however, it is unclear how this could have made the observed difference in SSQ scores between the simulated study and this study.

The answers to open-ended questions reveal interesting differences between the participants of this study and the simulated study, which could partly explain the differences in SSQ scores. Whereas in the simulated study only one person commented annoyance towards having to rotate the chair, here 4 people gave that comment (the participants saying "view follows" (\textit{``I did not need to turn when the robot turned. It was easy for me."}) and ``less self-rotations" were different participants). Moreover, it is surprising to observe three people consider CR less sickening, but these answers did not reveal any details of why. Thus, for comfort, it seems that simply the participants had different preferences when it comes to self-rotation than in the simulated study. This does not explain the non-significant difference in SSQ scores; however, as there is partial evidence from these studies together pointing towards unwinding reducing VR sickness, and the confirmed preference from both studies, we can conclude that unwinding does have clear advantages over coupled rotations also in a real-world telepresence scenario. Although in a real-world deployment, it may be beneficial to allow users to choose which condition to use, since it seems personal preferences play a role in this choice. 

The large difference in forced-choice answers to presence ("thinking back to both of the experiences, which one gave a better sense of being in the robot’s location?" 24 for UR, 8 for CR, p=.008) was a surprise: the simulated study \cite{suomalainen2022unwinding} showed promise towards this direction, which prompted us to explicitly measure it. Again, this was hinted directly in a few open-ended answers, such as \textit{``felt like I was more in charge of the situation/ what I was seeing and not just a passive observer"}. This result is in agreement with the theory of presence, since \textit{sensorimotor contingencies}, meaning that the user's physical actions are matched visually in the user's HMD view, are shown to increase the feeling of presence \cite{slater_psi:2009}. Whereas the nonsignificant difference in the SUS scores limits overemphasizing this result, remembering that the increase in presence was the main reason for using an \ac{hmd} in the first place, even a small increase with other positive effects is welcome. 

When not comparing the conditions, the fact that 11 out of 32 participants responded to greetings in the video is a positive sign of presence; the responses were despite the fact that we told the participants they are watching a recording, which also prompted a few participants to mention afterwards that they felt silly having responded. As this kind of behavior is exactly the reason to use an \ac{hmd} for this task, we can say that the results are still very promising. The next step is making live telepresence comparison between \ac{hmd}-based telepresence and conventional telepresence. 

Finally, after this study, we can confirm that the suggestions made for the robot's speed, and distances to people and objects in the simulation study, can be confirmed. The minimal distance to inanimate objects in this study was almost equal when compared against the simulation study ($0.95m$ and $0.9m$), and the result was the same, median score 4 in a 7-point Likert scale. The distance to people was slightly more in this study ($1.2m$ and $1m$), and this was reflected in the results; whereas in the simulation study the person was considered to be slightly too close at $1m$ (Median=3), in this study with $1.2m$ was considered suitable. Whereas this is not an exact comparison, we can conclude that doing telepresence research in VR is a good substitute, at least considering how the person aboard the robot sees the world and its distances.   

%\vspace{-1mm}
%\subsection{Limitations and Future Work}

%One of the SUS questions wasn't really well prepared to handle the case where users actually have been in the same location. 

%We have answered to bring the simulation study in our prior work \cite{suomalainen2022unwinding} to the real robot telepresence. Even though the results are statistically different from the simulation due to several aspects. We might want to observe within the robot and camera, our current robot movement makes the camera vibrate and it causes a 360 camera to jitter. Moreover, inspired by the participants’ reactions, we might consider bringing our system to live streaming, however there are inherent issues that include synchronization, latency, and bandwidth.

\section{CONCLUSIONS}\label{sec:con}
In this study, we evaluated whether the results shown for a simulated telepresence robot in \cite{suomalainen2022unwinding} also hold true for a telepresence robot with a 360\textdegree~camera; namely, that unwinding the rotations of the user, such that the user never sees rotational motion but must rotate themselves along with the robot, are preferred, more comfortable, and induce less VR sickness. However, this study found the same results only for preference as there was no statistically significant difference between the conditions in sickness or comfort. We believe the most likely reason for this difference was simply a different participant population, since there were several such hints in the open-ended questions. A slightly less likely explanation is that vibrations in the video feed caused by the robot motion increased the sickness scores, but this should have affected sickness in both conditions. We do believe, however, that this evidence shows the promise of unwinding, and also promise of virtual environments as a research platform for immersive telepresence. Future work will include a live immersive telepresence scenario, as well as testing unwinding on different, semi-autonomous steering methods. In the long-term, future research to enable two-way communication is merited as \ac{hmd}-based telepresence robots should be able to present the remote user's face to the people around the robot, similarly as current conventional telepresence robots \cite{kristoffersson2013review} do. There is already research towards showing a person's face who is wearing an HMD \cite{matsuda2021reverse}; thus, one should bring this technology to HMD-based telepresence robots to fully unlock the power of immersive telepresence.

%\section*{ACKNOWLEDGMENTS}

%%%%%%%%%%%%%%%%%%%%%%%%%%%%%%%%%%%%%%%%%%%%%%%%%%%%%%%%%%%%%%%%%%%%%%%%%%%%%%%%

\bibliographystyle{ieeetr}
\bibliography{references}

%\begin{thebibliography}{99}
%\end{thebibliography}

\end{document}